\documentclass[10pt,twocolumn,a4paper]{article}
\usepackage[top=2.2cm,bottom=2.2cm,left=1.8cm,right=1.8cm,columnsep=0.6cm]{geometry}
\usepackage{amsmath,amssymb,mathtools}
\usepackage{mathptmx}
\usepackage{courier}
\usepackage[T1]{fontenc}
\usepackage{graphicx}
\usepackage{booktabs}
\usepackage[numbers,sort&compress]{natbib}
\usepackage[font=small,labelfont=bf,labelsep=period]{caption}
\usepackage{microtype}
\usepackage{xcolor}
\usepackage[colorlinks=true,linkcolor=blue!50!black,citecolor=blue!50!black,urlcolor=blue!50!black]{hyperref}
\usepackage{amsthm}
\usepackage{titlesec}
\titleformat{\section}{\normalfont\bfseries\small\centering}{\Roman{section}.}{0.5em}{\MakeUppercase}
\titleformat{\subsection}{\normalfont\bfseries\small\centering}{\Alph{subsection}.}{0.5em}{}
\renewcommand{\thesection}{\Roman{section}}

\newtheorem{proposition}{Proposition}
\newtheorem{corollary}{Corollary}
\newcommand{\ud}{\mathrm{d}}
\newcommand{\Mirr}{M_{\rm irr}}

\begin{document}
\twocolumn[{
\begin{center}
{\Large\bfseries Penrose process in radiating Kerr--Vaidya black holes:\\[2pt] efficiency, energy transport, and the viability boundary\par}
\vspace{10pt}
{\normalsize Fabio Buffoli}\\[2pt]
{\small\itshape Universit\`a degli Studi di Brescia, Brescia, Italy}\\
{\small\texttt{fabio.buffoli@unibs.it}}
\vspace{10pt}
\begin{minipage}{0.84\textwidth}
\small
We study single-particle Penrose extraction from a rotating black hole whose mass decreases in time, described by the ingoing Kerr--Vaidya metric, imposing exact four-momentum conservation at the decay. We prove that the local efficiency is bounded by the instantaneous Kerr value, through a covariant bound that reduces to the Bardeen--Press--Teukolsky limit at the horizon, and we derive an exact transport law, $\ud E/\ud\tau=-\dot m\,(r/\Sigma)\,p_r^2$ for fixed spin parameter, by which every timelike geodesic gains energy while the hole loses mass. The net energy gain therefore splits into a local Penrose term and two transport terms, which we compute numerically as functions of spin and mass-loss rate for two mass-loss laws. With fixed spin parameter, extraction is possible only below a critical rate $\alpha_{\rm crit}(a)\simeq(M-a)/\Delta v$, set by the approach of the hole to extremality; with fixed $a/m$ no such boundary exists.
\end{minipage}
\vspace{14pt}
\end{center}
}]

\section{Introduction}
\label{sec:intro}

A Kerr black hole of mass $M$ and spin $a$ stores the rotational energy $M-\Mirr$, up to $29\%$ of its mass at extremality~\citep{christodoulou1970,bardeen1972}. Penrose showed that part of it can be extracted by particles~\citep{penrose1969,penrose1971}: a particle decays inside the ergosphere into a fragment of negative Killing energy, which falls into the hole, and a fragment that escapes with more energy than its parent. For a decay into two photons at the horizon of a hole with $a=M$, with the parent falling from rest at infinity, the efficiency is $(\sqrt2-1)/2\simeq20.7\%$~\citep{bardeen1972,wald1974,chandrasekhar1983}. Collisions near the horizon can exceed this value~\citep{piran1977,banados2009,harada2011,bejger2012,schnittman2014}, as can charged and magnetised variants~\citep{wagh1989,feiteira2024,zaslavskii2024}; the Blandford--Znajek mechanism~\citep{blandford1977} is the electromagnetic counterpart and the leading candidate for powering relativistic jets.

These results assume a stationary hole. A black hole that loses mass, through evaporation~\citep{hawking1975} or in any dynamical phase, is described in the simplest analytic form by the Vaidya metric~\citep{vaidya1951} and its rotating generalisations~\citep{carmeli1977,dahal2020,dahal2025}. In such spacetimes the Killing energy is no longer conserved and the ergosphere moves while a particle crosses it. Vertogradov~\citep{vertogradov2023} showed that in the charged Vaidya metric the generalised ergosphere exists only for a finite time. Bini et~al.~\citep{bini2011} showed that test particles in the Vaidya metric change energy as the mass changes. How these effects combine in the Penrose process of a rotating, radiating hole has not been determined.

In this paper we solve the problem for the equatorial Penrose decay in the Kerr--Vaidya metric, with exact four-momentum conservation. Our results are:
\begin{enumerate}
\item A covariant bound on the energy of the negative-energy fragment (Proposition~\ref{prop:bound}), valid for arbitrary fragment masses and parent energy. The decay must lie inside the instantaneous ergosphere, and the local efficiency cannot exceed the instantaneous Kerr value.
\item An exact energy-transport law (Proposition~\ref{prop:transport}): with fixed $a$, $\ud E/\ud\tau=-\dot m\,(r/\Sigma)\,p_r^2$ along every geodesic, so every timelike geodesic gains energy while the hole loses mass. For fixed $a/m$ an additional term of indefinite sign appears. The law follows from the Kerr--Schild structure of the metric and holds also for charged evaporating holes (Corollary~\ref{cor:ks}).
\item The energy budget of the process, $G=\Delta E_{\rm in}+(-E_2)+\Delta E_{\rm out}$, computed for spins $0.75\le a/M\le0.998$, mass-loss rates $10^{-4}\le\alpha\le10^{-1}$ and two mass-loss laws.
\item The viability boundary: with fixed $a$, Penrose extraction is possible only for $\alpha<\alpha_{\rm crit}(a)\simeq(M-a)/\Delta v$, where $\Delta v$ is the advanced time the particle needs to reach the ergosphere after the mass loss begins. The boundary is set by extremality, decreases with spin, and is absent when $a/m$ is fixed.
\end{enumerate}
Section~\ref{sec:metric} introduces the spacetime, Sec.~\ref{sec:analytic} the analytic results, Sec.~\ref{sec:numerics} the numerical method, Sec.~\ref{sec:results} the results, and Sec.~\ref{sec:discussion} their implications. We use $G=c=1$, the initial mass $M=1$, and energies per unit rest mass of the incoming particle.

\section{Kerr--Vaidya spacetime}
\label{sec:metric}

\subsection{Metric}
In ingoing coordinates $(v,r,\theta,\phi)$ the Kerr--Vaidya metric is~\citep{carmeli1977,dahal2020}
\begin{align}
\ud s^2 ={}& -\Bigl(1-\frac{2mr}{\Sigma}\Bigr)\ud v^2 + 2\,\ud v\,\ud r - 2a\sin^2\!\theta\,\ud r\,\ud\phi \notag\\
& - \frac{4amr\sin^2\!\theta}{\Sigma}\,\ud v\,\ud\phi + \Sigma\,\ud\theta^2 \notag\\
& + \frac{(r^2+a^2)^2-\Delta a^2\sin^2\!\theta}{\Sigma}\sin^2\!\theta\,\ud\phi^2,
\label{eq:metric}
\end{align}
with $\Sigma=r^2+a^2\cos^2\theta$, $\Delta=r^2-2mr+a^2$ and $m=m(v)$. We take $a\ge0$, so that the hole rotates in the direction of increasing $\phi$ and orbits with $L>0$ are prograde. For constant $m$ this is the Kerr metric in ingoing Kerr coordinates~\citep{misner1973}, and for $a=0$ it reduces to the ingoing Vaidya metric~\citep{vaidya1951}. Since the metric is invariant under $\theta\to\pi-\theta$, the equatorial plane $\theta=\pi/2$ is invariant under geodesic motion (initial data with $p_\theta=0$ there keep $p_\theta=0$), and we restrict to it. Inverting~\eqref{eq:metric} at $\theta=\pi/2$ gives the inverse metric in the $(v,r,\phi)$ block,
\begin{equation}
\begin{gathered}
g^{vv}=\frac{a^2}{r^2},\quad g^{vr}=\frac{r^2+a^2}{r^2},\quad g^{v\phi}=g^{r\phi}=\frac{a}{r^2},\\
g^{rr}=\frac{\Delta}{r^2},\quad g^{\phi\phi}=\frac{1}{r^2},
\end{gathered}
\label{eq:ginv}
\end{equation}
and only $g^{rr}$ depends on $m$, a property that makes the transport law of Sec.~\ref{subsec:transport} take a simple form.

The roots $r_\pm(v)=m\pm\sqrt{m^2-a^2}$ of $\Delta$ are the horizon radii of the instantaneous Kerr metric. For $\dot m<0$ the apparent horizon lies slightly outside $r_+(v)$~\citep{dahal2025}; we never need its exact location. The vector $l^\mu=-\delta^\mu_r$, with $l_\mu\ud x^\mu=-\ud v+a\sin^2\theta\,\ud\phi$, is null and future directed; it generates the ingoing principal null rays, along which $v$ is constant. Where $\Delta<0$, $g^{rr}=g^{\mu\nu}\partial_\mu r\,\partial_\nu r<0$, so $\nabla r$ is timelike; since $l^\mu\partial_\mu r=-1<0$, $r$ then decreases along every future-directed causal curve. The instantaneous ergosphere is the region $g_{vv}=2mr/\Sigma-1>0$, i.e.\ $r<2m(v)$ on the equator.

For constant $a$ the energy-momentum tensor obtained from~\eqref{eq:metric} through the Einstein equations is of Hawking--Ellis~\citep{hawking1973} type III, $T_{\mu\nu}=T_{00}k_\mu k_\nu+q_\mu k_\nu+q_\nu k_\mu$, with $k$ null and aligned with $l$~\citep{dahal2025}. For $\dot m<0$ it describes an ingoing flux of negative energy and violates the null energy condition. This is the standard Vaidya-type model of an evaporating hole~\citep{hiscock1981,dahal2020,dahal2025}.

\subsection{Mass-loss laws}
\label{subsec:laws}
The hole loses a mass $\delta m$ at constant rate $\alpha=-\dot m$ during a finite episode,
\begin{equation}
m(v)=\begin{cases} M, & v\le v_{\rm on},\\ M-\alpha(v-v_{\rm on}), & v_{\rm on}<v<v_{\rm end},\\ M-\delta m, & v\geq v_{\rm end},\end{cases}
\label{eq:mass}
\end{equation}
where $v_{\rm on}$ is the advanced time at which the loss begins and $v_{\rm end}=v_{\rm on}+\delta m/\alpha$. Outside the episode the spacetime is exactly Kerr, and $E=-p_v$ is a conserved Killing energy there. We consider two laws for the spin parameter:
\begin{itemize}
\item \textbf{Law A} (fixed $a$)~\citep{carmeli1977}. The dimensionless spin $a/m$ grows as the hole loses mass, and $\Delta$ has real roots only while $m\ge a$.
\item \textbf{Law B} (fixed $\chi=a/m$): Eq.~\eqref{eq:metric} with $a\to\chi\,m(v)$ in every component, a generalised Kerr--Vaidya metric with variable $a$~\citep{dahal2020,dahal2025}. The geometry shrinks self-similarly and never approaches extremality. Its energy-momentum tensor is of type IV~\citep{dahal2025}. We use law B to separate the dynamical effects of mass loss from the drift of $a/m$ that law A produces.
\end{itemize}

\section{Penrose process in a time-dependent background}
\label{sec:analytic}

\subsection{Local kinematics and efficiency bound}
\label{subsec:locality}
At a decay event $x_s$, four-momentum conservation $p_1=p_2+p_3$ and the mass shells $g^{\mu\nu}(x_s)p_{i\mu}p_{i\nu}=-\mu_i^2$ involve only $g_{\mu\nu}(x_s)$. This depends on $m(v_s)$ and $a(v_s)$ but not on their derivatives, which enter the connection. The kinematics of the decay is therefore that of Kerr with mass $m(v_s)$ and spin $a(v_s)$, and the content of this statement is the following bound.

\begin{proposition}[Decay bound]
\label{prop:bound}
A particle of unit rest mass and four-velocity $u$ decays at $x_s$ into fragments of rest masses $\mu_2,\mu_3$. In the parent rest frame fragment~2 has energy $\varepsilon_2$ and momentum $k$ ($\varepsilon_2^2-k^2=\mu_2^2$) along a unit vector $n\perp u$. Let $\xi=\partial_v$, which is the time-translation Killing field outside the episode, and define the energies $E_i=-p_i\cdot\xi=-p_{i\,v}$, so that $E_1=-u\cdot\xi$. Then
\begin{equation}
E_2=\varepsilon_2E_1-k\,n\cdot\xi\;\geq\;\varepsilon_2E_1-k\sqrt{E_1^2+g_{vv}(x_s)},
\label{eq:bound}
\end{equation}
with equality when $n$ is the unit projection of $\xi$ orthogonal to $u$.
\end{proposition}
\begin{proof}
In the parent rest frame $p_2=\varepsilon_2u+k\,n$, so $E_2=-p_2\cdot\xi=\varepsilon_2E_1-k\,n\cdot\xi$. The projection $\xi_\perp=\xi+(u\cdot\xi)u$ is orthogonal to $u$ and $\xi_\perp\cdot\xi_\perp=g_{vv}+E_1^2\ge0$. Since $n\perp u$, $n\cdot\xi=n\cdot\xi_\perp\le|\xi_\perp|$ by the Cauchy--Schwarz inequality on the spacelike subspace orthogonal to $u$.
\end{proof}

\begin{corollary}
\label{cor:ergo}
Since $k\le\varepsilon_2$ (because $\mu_2\ge0$) and $E_1>0$, Eq.~\eqref{eq:bound} allows $E_2<0$ only if $\sqrt{E_1^2+g_{vv}}>E_1$, i.e.\ $g_{vv}(x_s)>0$: the decay must lie inside the instantaneous ergosphere. For two equal-mass fragments emitted with rest-frame speed $\beta$ ($\varepsilon_2=1/2$, $k=\beta/2$),
\begin{equation}
\eta_{\rm loc}\equiv-\frac{E_2}{E_1}\le\eta_{\max}\equiv\tfrac12\Bigl(\beta\sqrt{1+g_{vv}/E_1^2}-1\Bigr).
\label{eq:etaloc}
\end{equation}
On the equator $g_{vv}=2m/r-1$; outside $r_+(v_s)$ the bound is largest at $r=r_+(v_s)$, where for $E_1=\beta=1$ it equals the Bardeen--Press--Teukolsky efficiency $\tfrac12(\sqrt{2m/r_+}-1)$~\citep{bardeen1972} at the instantaneous mass and spin.
\end{corollary}

Equation~\eqref{eq:etaloc} decreases with $E_1$, and so does the attainable $-E_2=\tfrac12(\beta\sqrt{E_1^2+g_{vv}}-E_1)$: a parent that reaches the decay with more energy produces a less negative $E_2$. Full four-momentum conservation fixes all fragment momenta, so no further constraint such as the area theorem has to be imposed; in Kerr a test particle crossing the horizon automatically satisfies $E_2-\Omega_HL_2\ge0$, with $\Omega_H$ the horizon angular velocity, and cannot decrease the irreducible mass~\citep{christodoulou1970,wald1974}. Conservation of the radial momentum is essential: if only $E$ and $L$ are conserved and the radial momentum of the escaping fragment is left free, the infalling fragment can carry up to the full rotational energy $m-\Mirr(m,a)$ of the hole, which no single test-particle event can remove.

\subsection{Energy-transport law}
\label{subsec:transport}
Geodesics are the integral curves of the Hamiltonian $H=\tfrac12g^{\mu\nu}p_\mu p_\nu$, with $\tau$ the affine parameter (proper time for massive particles) and $H=-\mu^2/2$. Hamilton's equation for $p_v$ gives $\ud p_v/\ud\tau=-\partial_vH=\tfrac12(\partial_vg_{\mu\nu})p^\mu p^\nu$, where $\partial_v$ acts only through $m(v)$ and $a(v)$; $p_\phi=L$ is conserved because no metric component depends on $\phi$. The energy $E=-p_v$ is therefore conserved only outside the episode.

\begin{proposition}[Energy transport]
\label{prop:transport}
For law A, along any geodesic of~\eqref{eq:metric},
\begin{equation}
\frac{\ud E}{\ud\tau}=-\dot m\,\frac{r}{\Sigma}\,(l\cdot p)^2=-\dot m\,\frac{r}{\Sigma}\,p_r^2 .
\label{eq:transportA}
\end{equation}
For law B, on the equator (where we need it),
\begin{equation}
\frac{\ud E}{\ud\tau}=-\dot m\Bigl[\frac{p_r^2}{r}-\chi\,(p_v+p_r)\,p^\phi\Bigr].
\label{eq:transportB}
\end{equation}
\end{proposition}
\begin{proof}
$\partial_mg_{\mu\nu}p^\mu p^\nu=(2r/\Sigma)(p^v-a\sin^2\theta\,p^\phi)^2=(2r/\Sigma)(l\cdot p)^2$ for all $\theta$, and $l\cdot p=-p_r$ since $l^\mu=-\delta^\mu_r$. On the equator $\partial_ag_{\mu\nu}p^\mu p^\nu=-2(p_v+p_r)\,p^\phi$ with $p^\phi=(ap_v+ap_r+p_\phi)/r^2$; law B has $\dot a=\chi\dot m$.
\end{proof}

The structure behind Eq.~\eqref{eq:transportA} is general. For constant $a$, the metric~\eqref{eq:metric} has the Kerr--Schild form $g_{\mu\nu}=\bar g_{\mu\nu}+F\,l_\mu l_\nu$, where $F=2mr/\Sigma$ and neither the background $\bar g_{\mu\nu}$ nor $l_\mu$ depends on $v$ (we verified symbolically that $g_{\mu\nu}-F\,l_\mu l_\nu$ is independent of $m$).

\begin{corollary}
\label{cor:ks}
For any metric $g_{\mu\nu}=\bar g_{\mu\nu}+F(v,r,\theta)\,l_\mu l_\nu$ with $v$-independent $\bar g_{\mu\nu}$ and $l_\mu$, geodesics obey $\ud E/\ud\tau=-\tfrac12(\partial_vF)\,(l\cdot p)^2$. In particular, for the Kerr--Newman--Vaidya metric, obtained from~\eqref{eq:metric} with $2mr\to2mr-Q^2$ and $\Delta=r^2-2mr+a^2+Q^2$, for which $l^\mu=-\delta^\mu_r$ is unchanged,
\begin{equation}
\frac{\ud E}{\ud\tau}=-\frac{\dot m\,r-Q\dot Q}{\Sigma}\,p_r^2 .
\label{eq:transportKN}
\end{equation}
\end{corollary}
\begin{proof}
$\ud p_v/\ud\tau=\tfrac12(\partial_vg_{\mu\nu})p^\mu p^\nu=\tfrac12(\partial_vF)(l\cdot p)^2$, and $l\cdot p=-p_r$; for Kerr--Newman--Vaidya $F=(2mr-Q^2)/\Sigma$. We verified symbolically, for all $\theta$, that $g_{\mu\nu}-F\,l_\mu l_\nu$ is independent of both $m$ and $Q$, that $l$ is null with $l^\mu=-\delta^\mu_r$, and that $\partial_mg_{\mu\nu}p^\mu p^\nu=(2r/\Sigma)\,p_r^2$ and $\partial_Qg_{\mu\nu}p^\mu p^\nu=-(2Q/\Sigma)\,p_r^2$.
\end{proof}

Equation~\eqref{eq:transportKN} applies to neutral particles; the electromagnetic force on charged particles adds a separate term. Loss of mass raises the energy of every geodesic, while loss of charge ($Q\dot Q<0$) lowers it, because it removes the repulsive $Q^2$ contribution to the potential. The same corollary covers Vaidya-type metrics with a radius-dependent mass function $m(v,r)$, used to model regular evaporating black holes.

Three consequences of Proposition~\ref{prop:transport} follow. First, for law A and $\dot m<0$ the energy grows monotonically along every timelike geodesic and every null geodesic except the ingoing principal null rays, for which $p_r=0$. The effect is present for $a=0$, where Eq.~\eqref{eq:transportA} reads $\ud p_v/\ud\tau=(\dot m/r)\,p_r^2$; this is the ingoing counterpart of the result of Bini et~al.~\citep{bini2011} for the outgoing Vaidya metric, $\ud p_u/\ud\tau=(\partial_uM/r)\,p_r^2$, with the retarded time $u$ in place of $v$. Second, for slow mass loss
\begin{equation}
\Delta E\simeq\alpha\int\frac{r}{\Sigma}\,p_r^2\,\ud\tau,
\label{eq:firstorder}
\end{equation}
obtained by evaluating the right-hand side of~\eqref{eq:transportA} on the static geodesic with the same initial data, which is correct to first order in $\alpha$, and integrating over the part of the path inside the episode. For an impulsive episode one writes $\ud\tau=\ud v/u^v$ (with $u^v=\ud v/\ud\tau>0$ on the path), so that $\ud E/\ud m=-(r/\Sigma)\,p_r^2/u^v$ and $\Delta E\to\delta m\,[(r/\Sigma)p_r^2/u^v]_{v_{\rm on}}$, independent of $\alpha$. Third, for law B the second term of~\eqref{eq:transportB} is positive when $(E-p_r)\,p^\phi>0$ and can be negative for counter-rotating particles.

\subsection{Energy budget}
\label{subsec:budget}
A parent launched with Killing energy $E_0$ before the episode decays at $x_s$ with energy $E_{1s}$ into fragments with energies $E_{2s}$ and $E_{3s}$; fragment~3 escapes and is read out after the episode with Killing energy $E_\infty$. Using $E_{3s}=E_{1s}-E_{2s}$ (the $v$-component of $p_1=p_2+p_3$), the net gain is
\begin{equation}
G\equiv E_\infty-E_0=\underbrace{E_{1s}-E_0}_{\Delta E_{\rm in}}+\underbrace{(-E_{2s})}_{\text{local}}+\underbrace{E_\infty-E_{3s}}_{\Delta E_{\rm out}} .
\label{eq:budget}
\end{equation}
$G$ is invariant, because $E_0$ and $E_\infty$ are Killing energies. The three terms are unchanged by the coordinate transformations $v\to v+f(r)$ that preserve the form of~\eqref{eq:metric}, since $\partial_v$ at fixed $(r,\phi)$ is the same vector field; they are defined with respect to $\partial_v$, which coincides with the time-translation Killing field outside the episode, and Proposition~\ref{prop:bound} uses the same field. Because $E$ is conserved in static Kerr, $\Delta E_{\rm in}$ and $\Delta E_{\rm out}$ contain no gravitational binding energy: both vanish identically for $\alpha=0$, and both are given by integrals of~\eqref{eq:transportA}--\eqref{eq:transportB}.

\subsection{Viability boundary}
\label{subsec:extremality}
Under law A with an uncapped episode, $\Delta$ loses its real roots at
\begin{equation}
v_{\rm ext}=v_{\rm on}+\frac{M-a}{\alpha},
\label{eq:vext}
\end{equation}
after which $\Delta>0$ everywhere, $\nabla r$ is spacelike everywhere, and the region in which $r$ decreases along all future-directed causal curves disappears: the model no longer describes a black hole. A negative-energy decay requires the parent to be inside the instantaneous ergosphere (Corollary~\ref{cor:ergo}) and outside $r_+$ before $v_{\rm ext}$. If $\Delta v$ is the shortest advanced time between the start of the episode and the entry of the parent into the ergosphere, extraction is therefore possible only for
\begin{equation}
\alpha<\alpha_{\rm crit}(a)\simeq\frac{M-a}{\Delta v}.
\label{eq:alphacrit}
\end{equation}
The boundary decreases with spin as $M-a$ and closes at $a=M$. Under law B, $a/m$ is constant and no boundary exists.

\section{Numerical method}
\label{sec:numerics}

We integrate Hamilton's equations on the equator for $(v,r,\phi,p_v,p_r,p_\phi)$, using~\eqref{eq:ginv} and $\ud p_v/\ud\tau=-(\dot m\,\partial_mH+\dot a\,\partial_aH)$ with $\partial_mH=-p_r^2/r$ and $\partial_aH=(p_v+p_r)\,p^\phi$. The integrator is an adaptive Dormand--Prince 5(4) scheme~\citep{dormand1980} with relative and absolute tolerances $10^{-10}$ and $10^{-12}$, compiled with Numba~\citep{lam2015}, with root-bracketing event location. The analytic expressions used in the code (inverse metric, Hamilton's equations, Propositions~\ref{prop:bound} and~\ref{prop:transport}) were derived and checked symbolically with SymPy~\citep{meurer2017}.

\paragraph{Decay.} The parent decays into two fragments of equal rest mass emitted back to back with rest-frame speed $\beta$ along $n(\psi)=\cos\psi\,e_1+\sin\psi\,e_2$, where $e_1$ is the unit projection of $\partial_v$ orthogonal to $u$ and $e_2$ completes an orthonormal equatorial frame; $\psi=0$ saturates~\eqref{eq:bound}. Four-momentum is conserved to machine precision. Unless stated otherwise $\beta=1$ (two photons).

\paragraph{Capture and escape.} Fragment~2 is captured when it reaches $r<r_f-10^{-3}M$, with $r_f\equiv r_+(M-\delta m)$. Under both laws $r_+(v)\ge r_f$ and $r_-(v)<r_f$ for all our parameters, so the band $r_-(v)<r<r_f$ lies in $\Delta<0$ at all times and a fragment below $r_f$ can never return to larger radii. Fragment~3 escapes when it reaches $r>200M$ moving outwards with $v>v_{\rm end}$, so that its energy is a Killing energy.

\paragraph{Search.} The parent is launched from $r=10M$ with $E_0=1$ and angular momentum $L_0\in[-4,4.5]$ on a grid of 69 values. The episode starts when the parent crosses $r=3M$, so that it overlaps the plunge. The decay occurs at the first crossing of $r=r_+(v)+\delta$ inside the instantaneous ergosphere, for 14 values of $\delta\in[10^{-3},0.7]M$, and $\psi$ is scanned over 288 values; the best grid points are then refined continuously (Appendix~\ref{app:conv}). A decay is valid if $E_{2s}<0$, fragment~2 is captured and fragment~3 escapes; for each (law, $a_0$, $\alpha$) we report the valid decay with the largest $G$. We take $\delta m=0.05M$ for law B and $\delta m=\min[0.05,\,0.9(1-a_0)]M$ for law A. The factor $0.9$ keeps the final hole sub-extremal ($a/m_f\le0.9998$ for $a_0\le0.998$) and the capture band non-empty.

\paragraph{Validation.} Hamilton's equations agree with finite differences of $H$, and the coded $\ud E/\ud\tau$ with~\eqref{eq:transportA}--\eqref{eq:transportB}, to $7\times10^{-10}$ over $400$ random phase-space points. In static Kerr, $E$ is conserved exactly and $H$ to $2\times10^{-11}$; in dynamic runs $H$ is conserved to $2\times10^{-7}$ and $E(\tau)$ agrees with the integral of~\eqref{eq:transportA}--\eqref{eq:transportB} to $3\times10^{-6}$. For the Kerr--Newman--Vaidya extension we used an independent code, in which the equatorial Hamiltonian is built symbolically from the full metric with $m(v)$ and $Q(v)$ and integrated with a separate Dormand--Prince 8(5,3) scheme: for $a=0.9$, $Q_0=0.3$ and $\dot m,\dot Q\in\{0,\pm2\times10^{-3}\}$, $E(\tau)$ agrees with the integral of Eq.~\eqref{eq:transportKN} to $8\times10^{-10}$, and a discharge alone ($\dot m=0$, $Q\dot Q<0$) lowers the energy, as stated. For $a=0$, outgoing radial null rays obey $\ud r/\ud v=\tfrac12(1-2m(v)/r)$ exactly; the integrator reproduces an independent solution of this equation to $10^{-9}$ for $\alpha=10^{-3}$--$10^{-1}$. A photon decay at $r=r_+(1+10^{-7})$ with $E_1=1$ reproduces $\tfrac12(\sqrt{2M/r_+}-1)$ to six digits for $a=0.9$--$0.9999$. Tightening the tolerances by $10^2$ and moving the escape radius to $2000M$ changes the optimal decays by less than $10^{-6}$. Resolution tests are reported in Appendix~\ref{app:conv}.

\section{Results}
\label{sec:results}

\subsection{Static benchmark and spin dependence}
\label{subsec:static}
In stationary Kerr ($\alpha=0$) escaping photon decays with a negative-energy partner exist for $a_0\ge0.8$ in our launch family and are absent at $a_0=0.75$. The optimal gains are $G=0.011$, $0.030$, $0.055$, $0.092$, $0.129$, $0.150$ and $0.180$ for $a_0=0.8$, $0.85$, $0.9$, $0.95$, $0.98$, $0.99$ and $0.998$. At high spin they approach the horizon value $\tfrac12(\sqrt{2M/r_+}-1)$ ($93\%$ of it at $a_0=0.99$, $97\%$ at $0.998$); at lower spin they fall well below it, because a photon emitted along the optimal direction~\eqref{eq:bound} close to the horizon is recaptured. The optimal decays often follow parents on near-critical orbits that whirl around the hole before decaying. Massive fragments narrow the window: at $a_0=0.9$ valid decays exist for $\beta=0.95$ ($G=0.027$) and not for $\beta\le0.8$.

\begin{figure}[t]
\centering
\includegraphics[width=\columnwidth]{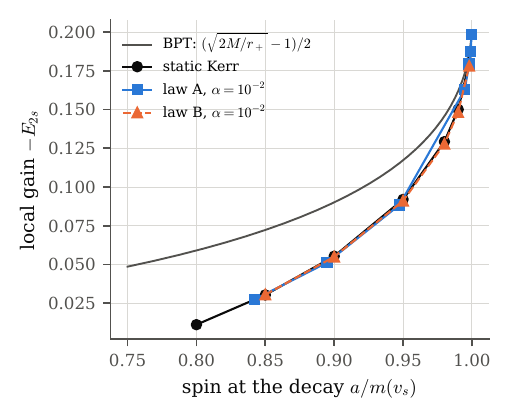}
\caption{Local term $-E_{2s}$ of the optimal escaping decay versus the dimensionless spin of the hole at the moment of the decay, $a/m(v_s)$, for static Kerr (circles) and for laws A (squares) and B (triangles) at $\alpha=10^{-2}$. Each point is one initial spin $a_0$; under law A the points move to larger $a/m(v_s)$ because the hole spins up as it loses mass. The grey curve is the bound~\eqref{eq:etaloc} at the horizon for $E_1=\beta=1$, i.e.\ the Bardeen--Press--Teukolsky efficiency.}
\label{fig:spin}
\end{figure}

Figure~\ref{fig:spin} shows the local term $-E_{2s}$ of the optimal decays as a function of the spin at the decay, $a/m(v_s)$, for static Kerr and for both laws at $\alpha=10^{-2}$. The three sets lie on a single curve. The local gain is a function of the instantaneous spin, as Proposition~\ref{prop:bound} requires; the larger values reached under law A come entirely from the drift of $a/m(v_s)$ towards extremality.

\subsection{Pure transport: scattering orbits}
\label{subsec:flyby}
\begin{figure}[t]
\centering
\includegraphics[width=\columnwidth]{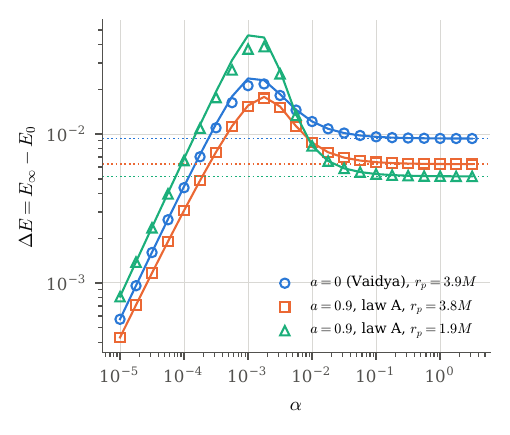}
\caption{Energy $\Delta E=E_\infty-E_0$ gained by scattering geodesics (no decay, $E_0=1.1$) as a function of the mass-loss rate, for an episode of $\delta m=0.05$ starting at $r=6M$. Three orbits are shown: $a=0$ (circles), and $a=0.9$ under law A with pericentre $r_p=3.8M$ (squares) and $1.9M$, inside the ergosphere (triangles). Symbols: full integration; solid lines: first-order prediction~\eqref{eq:firstorder}; dotted lines: impulsive limit $\delta m\,[(r/\Sigma)p_r^2/u^v]_{v_{\rm on}}$.}
\label{fig:flyby}
\end{figure}

To isolate transport we follow scattering geodesics without decay: $E_0=1.1$, launched from $r=50M$, with an episode of $\delta m=0.05$ starting when the particle crosses $r=6M$ inbound (Fig.~\ref{fig:flyby}). Three regimes appear. For slow loss $\Delta E\propto\alpha$, with $\Delta E/\alpha=40$--$81$, and Eq.~\eqref{eq:firstorder} reproduces the full integration to within $3\%$ for $\alpha\le10^{-4}$. $\Delta E$ peaks at $\alpha\simeq(1$--$2)\times10^{-3}$, where the episode duration $\delta m/\alpha\simeq30M$ matches the time the particle spends at small radii, with $\Delta E=0.022$ for $a=0$, $0.017$ for $a=0.9$ at pericentre $3.8M$, and $0.039$ for an orbit that enters the ergosphere. For fast loss $\Delta E$ tends to the impulsive value. The non-rotating hole produces a transport gain of the same size as the rotating one: transport is driven by the mass loss, not by the rotation. For $a=0$ the ergosphere coincides with the horizon, so the local term vanishes identically and the whole gain, up to $\Delta E=0.022$ here, is transport.

\subsection{Energy budget versus mass-loss rate}
\label{subsec:budget_results}
\begin{figure*}[t]
\centering
\includegraphics[width=0.92\textwidth]{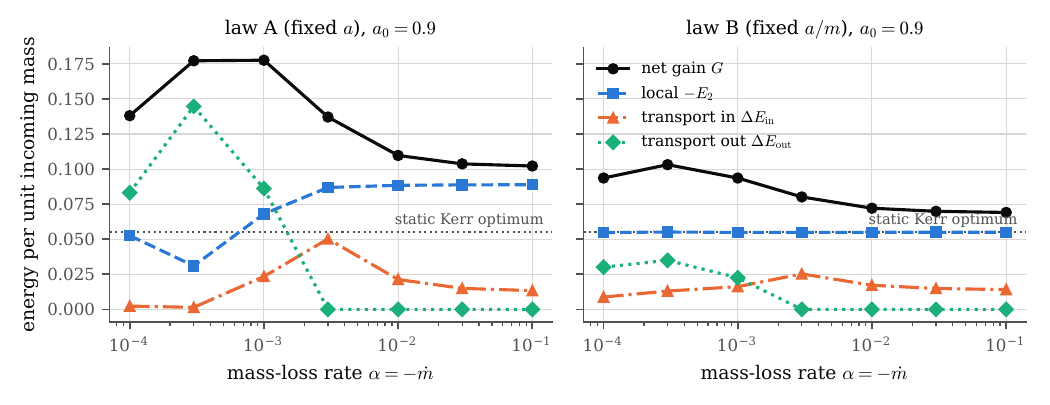}
\caption{Energy budget~\eqref{eq:budget} of the optimal escaping photon decay as a function of the mass-loss rate, for $a_0=0.9$ under law A (left) and law B (right): net gain $G$ (circles), local Penrose term $-E_{2s}$ (squares), and transport before ($\Delta E_{\rm in}$, triangles) and after ($\Delta E_{\rm out}$, diamonds) the decay. The dotted line is the optimum in static Kerr ($\alpha=0$). All energies are per unit rest mass of the parent.}
\label{fig:budget}
\end{figure*}

\begin{figure*}[t]
\centering
\includegraphics[width=0.92\textwidth]{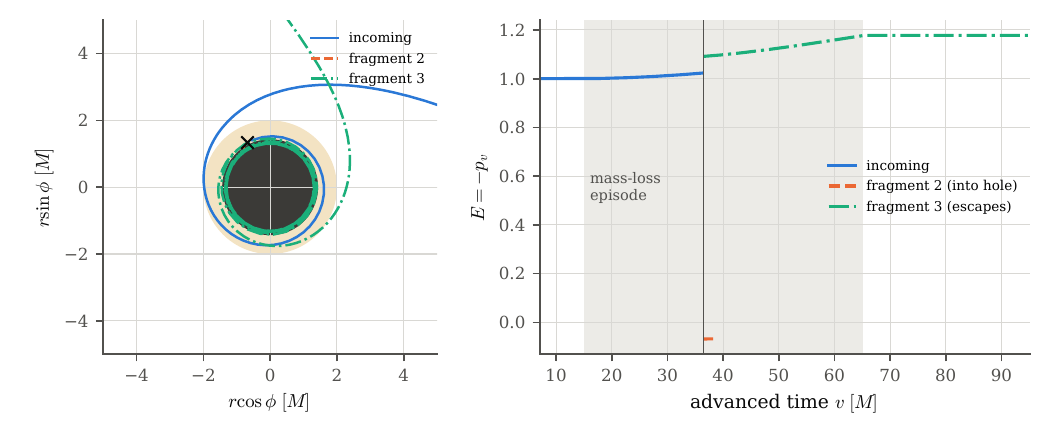}
\caption{Optimal decay for law A, $a_0=0.9$, $\alpha=10^{-3}$, $\delta m=0.05$. Left: equatorial orbits, plotted as $(r\cos\phi,r\sin\phi)$ with $\phi$ the ingoing azimuth of Eq.~\eqref{eq:metric}; the parent (solid) whirls around the hole before decaying, fragment~2 (dashed) falls in and fragment~3 (dash-dotted) escapes; the cross marks the decay, the shaded annulus the initial ergosphere and the dark disc the initial radius $r_+$. Right: $E=-p_v$ versus advanced time; $E$ changes only during the mass-loss episode (shaded) and jumps at the decay. The parent gains $\Delta E_{\rm in}=0.023$, the decay gives $-E_{2s}=0.068$, and the escaping fragment gains $\Delta E_{\rm out}=0.086$, for $G=0.178$.}
\label{fig:example}
\end{figure*}

Table~\ref{tab:budget} and Fig.~\ref{fig:budget} give the budget~\eqref{eq:budget} of the optimal decays; Fig.~\ref{fig:example} shows one of them. Three regimes appear as the episode duration $\delta m/\alpha$ is compared with the time the fragments spend near the hole.
\begin{itemize}
\item \emph{Fast mass loss} ($\alpha\gtrsim10^{-2}$): the episode ends before the decay, $\Delta E_{\rm out}=0$, and the decay takes place in the final Kerr hole. Under law A the local term grows with the final spin: at $a_0=0.95$ it reaches $0.163$ against $0.092$ in the static case, with $a/m(v_s)=0.995$. Under law B the local term equals its static value to within $2\%$ at every spin, and $G$ exceeds it only by $\Delta E_{\rm in}\simeq0.013$--$0.017$.
\item \emph{Intermediate rates} ($10^{-4}\lesssim\alpha\lesssim10^{-3}$): the escaping fragment is still inside the episode and $\Delta E_{\rm out}$ becomes the largest term. The optimal decays emit fragment~3 close to the boundary between capture and escape, so that it remains near the hole for $10^2$--$5\times10^2M$ while the hole keeps losing mass, and is released with a large gain. This horizon-release channel gives the largest net gains of the study, $G=0.33$ at $a_0=0.95$ under law A, more than three times the static optimum. These gains require fragment~3 to be emitted within a narrow range of directions and to reside near the hole for $10^2$--$5\times10^2M$; they are optimal, finely tuned configurations rather than generic outcomes, distinct local optima of comparable value exist (Appendix~\ref{app:conv}), and they vanish as $\alpha\to0$.
\item \emph{Slow mass loss} ($\alpha\to0$ at fixed $\delta m$): every transport term is proportional to $\alpha$ times the residence time of the particles near the hole, Eq.~\eqref{eq:firstorder}, and $G$ returns to its static value. The sizeable gains still present at $\alpha=10^{-4}$ reflect residence times of several $10^2M$, i.e.\ $\alpha\,\tau_{\rm res}\sim0.05$, not a breakdown of this limit. Conversely, for a short episode ($\delta m\to0$ at fixed $\alpha$) the transport terms are bounded by $\delta m$ times the maximum of $\ud E/\ud m$ along the path and vanish, and the local term tends to its static value.
\end{itemize}
The dependence on the size of the episode is simple. At $a_0=0.9$ and $\alpha=10^{-2}$, varying $\delta m$ from $0.01$ to $0.08$ leaves the local term of law B at $0.055$ while $\Delta E_{\rm in}$ grows from $0.003$ to $0.032$, nearly in proportion to $\delta m$; under law A the local term also grows, from $0.061$ to $0.122$, because the final spin does. At $\alpha=10^{-3}$, where fragment~3 is released during the episode, halving $\delta m$ lowers $G$ from $0.178$ to $0.098$.

\begin{table}[t]
\centering
\caption{Energy budget~\eqref{eq:budget} of the optimal decay (photon fragments) for three initial spins $a_0$ and the two mass-loss laws; rows with $\alpha=0$ are static Kerr and are common to both laws. Columns: spin at the decay $a/m(v_s)$, transport before the decay, local Penrose term, transport after the decay, and net gain $G$, which is their sum. $\delta m=0.05$ except for law A at $a_0=0.95$ ($\delta m=0.045$) and $a_0=0.99$ ($\delta m=0.009$); $a/m$ is the spin at the decay.}
\label{tab:budget}
\scriptsize
\setlength{\tabcolsep}{3.2pt}
\begin{tabular}{cccccccc}
\toprule
law & $a_0$ & $\alpha$ & $a/m$ & $\Delta E_{\rm in}$ & $-E_{2s}$ & $\Delta E_{\rm out}$ & $G$\\
\midrule
-- & 0.9 & 0 & 0.900 & 0.000 & 0.055 & 0.000 & 0.055 \\
A & 0.9 & $10^{-4}$ & 0.902 & 0.002 & 0.053 & 0.083 & 0.138 \\
A & 0.9 & $10^{-3}$ & 0.920 & 0.023 & 0.068 & 0.086 & 0.178 \\
A & 0.9 & $10^{-2}$ & 0.947 & 0.021 & 0.088 & 0.000 & 0.110 \\
A & 0.9 & $10^{-1}$ & 0.947 & 0.013 & 0.089 & 0.000 & 0.102 \\
-- & 0.95 & 0 & 0.950 & 0.000 & 0.092 & 0.000 & 0.092 \\
A & 0.95 & $10^{-4}$ & 0.951 & 0.001 & 0.076 & 0.248 & 0.325 \\
A & 0.95 & $10^{-3}$ & 0.970 & 0.028 & 0.125 & 0.132 & 0.285 \\
A & 0.95 & $10^{-2}$ & 0.995 & 0.017 & 0.163 & 0.000 & 0.180 \\
A & 0.95 & $10^{-1}$ & 0.995 & 0.010 & 0.164 & 0.000 & 0.174 \\
-- & 0.99 & 0 & 0.990 & 0.000 & 0.150 & 0.000 & 0.150 \\
A & 0.99 & $10^{-4}$ & 0.993 & 0.010 & 0.165 & 0.061 & 0.236 \\
A & 0.99 & $10^{-3}$ & 0.999 & 0.006 & 0.187 & 0.000 & 0.193 \\
A & 0.99 & $10^{-2}$ & 0.999 & 0.002 & 0.188 & 0.000 & 0.190 \\
A & 0.99 & $10^{-1}$ & 0.999 & 0.002 & 0.188 & 0.000 & 0.190 \\
\midrule
B & 0.9 & $10^{-4}$ & 0.900 & 0.009 & 0.055 & 0.030 & 0.094 \\
B & 0.9 & $10^{-3}$ & 0.900 & 0.016 & 0.055 & 0.023 & 0.094 \\
B & 0.9 & $10^{-2}$ & 0.900 & 0.017 & 0.055 & 0.000 & 0.072 \\
B & 0.9 & $10^{-1}$ & 0.900 & 0.014 & 0.055 & 0.000 & 0.069 \\
B & 0.95 & $10^{-4}$ & 0.950 & 0.005 & 0.091 & 0.041 & 0.136 \\
B & 0.95 & $10^{-3}$ & 0.950 & 0.016 & 0.091 & 0.022 & 0.129 \\
B & 0.95 & $10^{-2}$ & 0.950 & 0.017 & 0.091 & 0.000 & 0.108 \\
B & 0.95 & $10^{-1}$ & 0.950 & 0.014 & 0.091 & 0.000 & 0.105 \\
B & 0.99 & $10^{-4}$ & 0.990 & 0.003 & 0.148 & 0.055 & 0.206 \\
B & 0.99 & $10^{-3}$ & 0.990 & 0.015 & 0.149 & 0.019 & 0.183 \\
B & 0.99 & $10^{-2}$ & 0.990 & 0.017 & 0.148 & 0.000 & 0.165 \\
B & 0.99 & $10^{-1}$ & 0.990 & 0.013 & 0.149 & 0.000 & 0.162 \\
\bottomrule
\end{tabular}
\end{table}

\subsection{Local bound}
\label{subsec:locality_results}
\begin{figure}[t]
\centering
\includegraphics[width=\columnwidth]{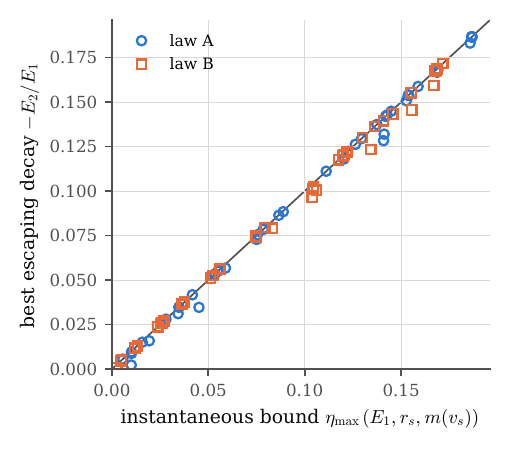}
\caption{Best local efficiency $\eta_{\rm loc}=-E_{2s}/E_{1s}$ among the escaping decays of each combination of law, spin and $\alpha$, versus the bound $\eta_{\max}$ of Eq.~\eqref{eq:etaloc} evaluated at the same event with the parent's actual $E_{1s}$ and the instantaneous $m(v_s)$. The line marks equality; no point lies above it.}
\label{fig:locality}
\end{figure}

Over all $100$ combinations of law, spin and $\alpha$ with valid decays ($9.0\times10^4$ valid decays in total), no decay exceeds the bound~\eqref{eq:etaloc} evaluated with its own $E_{1s}$ and $m(v_s)$; the largest excess is $6\times10^{-8}$, at the level of the integration tolerance (Fig.~\ref{fig:locality}). The best local efficiency saturates the bound, to within $10^{-4}$, in $66$ of the $100$ combinations, namely whenever the decay along $\psi=0$ lets fragment~3 escape; in the others the saturating fragment is recaptured and the best escaping decay lies below the bound.

\subsection{Viability boundary}
\label{subsec:viability}
\begin{figure}[t]
\centering
\includegraphics[width=\columnwidth]{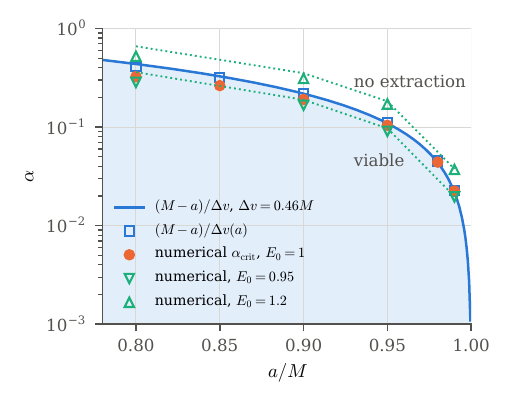}
\caption{Viability boundary of law A (uncapped linear mass loss). Filled circles: numerical $\alpha_{\rm crit}$ for parents launched with $E_0=1$; open squares: Eq.~\eqref{eq:alphacrit} with the spin-dependent $\Delta v$; solid line: Eq.~\eqref{eq:alphacrit} with the mean $\Delta v$. Open triangles and dotted lines: numerical boundary and Eq.~\eqref{eq:alphacrit} for $E_0=0.95$ (down) and $E_0=1.2$ (up). For $E_0=1$, negative-energy decays are possible only in the shaded region.}
\label{fig:viability}
\end{figure}

We determine $\alpha_{\rm crit}(a)$ for law A with an uncapped episode by bisection: a rate is viable if at least one trajectory of the launch family enters the instantaneous ergosphere, outside $r_+(v)$, before $v_{\rm ext}$. The static trajectories give $\Delta v=0.44$--$0.48M$ between the start of the episode at $r=3M$ and the entry into the ergosphere. Figure~\ref{fig:viability} compares the numerical boundary with Eq.~\eqref{eq:alphacrit}. We find $\alpha_{\rm crit}=0.32$, $0.26$, $0.19$, $0.10$, $0.044$ and $0.022$ for $a=0.8$, $0.85$, $0.9$, $0.95$, $0.98$ and $0.99$. Equation~\eqref{eq:alphacrit} agrees to $1.4\%$ at $a=0.99$ and overestimates the boundary by up to $27\%$ at $a=0.8$, where $\alpha_{\rm crit}$ is large and the ergosphere shrinks appreciably while the particle approaches it. The boundary decreases with spin and closes at $a=M$. $\Delta v$ varies by only $10\%$ across the spin range because the episode starts at the same radius for all spins; a different launch set-up changes $\Delta v$, and hence the prefactor of Eq.~\eqref{eq:alphacrit}, but not the $(M-a)$ scaling. We checked this by repeating the computation for parents launched with $E_0=0.95$ and $E_0=1.2$: $\Delta v$ changes to $0.51$--$0.55M$ and $0.27$--$0.30M$, $\alpha_{\rm crit}(0.99)$ to $0.020$ and $0.037$, and Eq.~\eqref{eq:alphacrit} with the corresponding $\Delta v$ reproduces the numerical boundary to within $1.4\%$ and $1.3\%$ at $a=0.99$ in both cases. The relation $\alpha_{\rm crit}\simeq(M-a)/\Delta v$ is therefore robust, while the value of $\Delta v$ is a property of the infalling matter.

\section{Discussion}
\label{sec:discussion}

\paragraph{Rotational energy and transport.} With respect to $\partial_v$, only the local term $-E_{2s}$ requires an ergosphere, and Proposition~\ref{prop:bound} bounds it at the instantaneous mass and spin. The transport terms exist for $a=0$ and are driven by the mass loss itself (Sec.~\ref{subsec:flyby}). The invariant gain $G$ therefore combines rotational extraction with transport, and the two must be separated before an enhancement of Penrose extraction can be attributed to the dynamics of the hole. Under law A a third effect, the growth of $a/m$, raises the local term (Fig.~\ref{fig:spin}); law B removes it.

\paragraph{Astrophysical scales.} All dynamical corrections scale as $\alpha$ times the residence time of the particles near the hole, $\sim10$--$10^3M$. For Hawking evaporation, in the simplest blackbody estimate, $\alpha\simeq(m_{\rm P}/m)^2/(15360\pi)$, of order $10^{-81}$ for a solar-mass hole and $10^{-44}$ for a $10^{12}\,$kg primordial hole~\citep{hawking1975}. With $\Delta E/\alpha\simeq40$--$80$ for single passages (Sec.~\ref{subsec:flyby}) and up to $\sim10^3$ for the longest residence times found in Sec.~\ref{subsec:budget_results}, the transport correction for an evaporating hole is $\Delta E\lesssim2\times10^{-2}\,(m_{\rm P}/m)^2$ per unit rest mass. It reaches $1\%$ only for $m\lesssim1.4\,m_{\rm P}$, where the semiclassical description of evaporation no longer applies. Single-particle Penrose extraction from an evaporating black hole, including a primordial one in its final stage, is therefore extraction from Kerr at the instantaneous mass and spin down to the Planck scale, with corrections given by Eq.~\eqref{eq:firstorder}. The regime $\alpha\gtrsim10^{-4}$ explored above, where the null energy condition is strongly violated, is where the transport terms and the viability boundary become visible; it characterises the Kerr--Vaidya model itself.

\paragraph{Scope.} We considered equatorial motion, test particles and equal-mass decays. Proposition~\ref{prop:bound} holds for any decay at any $\theta$, and Proposition~\ref{prop:transport} (for law A at all $\theta$) holds for any geodesic, so both apply unchanged to off-equatorial motion; the transport law applies also to the particles entering collisional Penrose processes. Penrose decays of charged particles in the Kerr--Newman--Vaidya family, where Eq.~\eqref{eq:transportKN} describes the neutral fragments, are a natural extension.

\section{Conclusions}
\label{sec:conclusions}

We have studied the Penrose process in the Kerr--Vaidya spacetime with exact four-momentum conservation. The local efficiency is bounded by the instantaneous Kerr value through the covariant bound~\eqref{eq:bound}, which requires the decay to lie inside the instantaneous ergosphere and reduces to the Bardeen--Press--Teukolsky efficiency at the horizon; $9\times10^4$ numerically integrated decays respect it. The energy of every geodesic obeys the exact transport law~\eqref{eq:transportA}--\eqref{eq:transportB}, which follows from the Kerr--Schild structure of the metric and extends to the Kerr--Newman--Vaidya family, Eq.~\eqref{eq:transportKN}, so the net gain consists of a local Penrose term and two transport terms. The local term depends only on the spin at the decay. Transport dominates for mass-loss rates $10^{-4}\lesssim\alpha\lesssim10^{-3}$ through the release of fragments trapped near the shrinking hole, and vanishes linearly as $\alpha\to0$. With fixed $a$, extraction is possible only below $\alpha_{\rm crit}(a)\simeq(M-a)/\Delta v$, a boundary set by the approach to extremality, which decreases with spin and is absent when $a/m$ is fixed.

\section*{Data availability}
The code (symbolic derivations, Numba integrator, validation suite, parameter studies) and the data behind all figures and tables are available at \url{https://github.com/FabioB95/kerr-vaidya-penrose-optimizer}.

\appendix
\renewcommand{\thesection}{\Alph{section}}
\titleformat{\section}{\normalfont\bfseries\small\centering}{APPENDIX \Alph{section}:}{0.5em}{\MakeUppercase}
\section{Convergence}
\label{app:conv}
The optimal decays are obtained by a grid search followed by Nelder--Mead refinement~\citep{nelder1965} in $(L_0,\ln\delta,\psi)$ from the five best grid points. Repeating the search with a grid of twice the resolution in $L_0$ and $\psi$ and 27 values of $\delta$ gives the same optimal $G$ to four digits for law A at $(a_0,\alpha)=(0.9,0)$, $(0.9,10^{-2})$, $(0.99,0)$ and for law B at $(0.99,10^{-3})$, and to $0.3\%$ for law A at $(0.9,10^{-3})$. For the horizon-release optima, which are finely tuned, the refined search on the finer grid converges to a different local optimum: at $(a_0,\alpha)=(0.95,3\times10^{-4})$ it finds $G=0.311$ against $0.326$. Every reported $G$ is attained by an explicit valid decay, so the values in Table~\ref{tab:budget} are lower bounds on the true optimum within our set-up, exact up to the integration tolerance for each configuration; for the finely tuned cases, distinct local optima differ by up to $5\%$. The grid search alone is not sufficient at high spin, where the optimum lies in a narrow window of $\delta$: for $a_0=0.99$ the grid optimum rises from $0.122$ to $0.143$ when the grid is refined, while the refined optimum is $0.150$ in both cases.

\bibliographystyle{unsrtnat}
\bibliography{references}
\end{document}